\documentclass[preprint,12pt,times,number,sort&compress]{elsarticle}

\usepackage{amssymb}
\usepackage{amsmath}
\usepackage{ptdr-definitions}
\usepackage{heppennames2}
\usepackage{lineno}

\journal{Subatomic Particles and Cosmology.}

\newcommand{\Hcc}{\ensuremath{\PH{}\rightarrow \PQc \PAQc}\xspace}
\newcommand{\Hbb}{\ensuremath{\PH{}\rightarrow \PQb \PAQb}\xspace}
\newcommand{\Zcc}{\ensuremath{\PZ{}\rightarrow \PQc \PAQc}\xspace}
\newcommand{\Zbb}{\ensuremath{\PZ{}\rightarrow \PQb \PAQb}\xspace}

\newcommand{\ttH}{\ttbar{}\PH{}\xspace}
\newcommand{\ttZ}{\ttbar{}\PZ{}\xspace}
\newcommand{\ttX}{\ttbar{}\PX{}\xspace}

\newcommand{\ttHbb}{\ttH{}(\Hbb{})\xspace}
\newcommand{\ttHcc}{\ttH{}(\Hcc{})\xspace}
\newcommand{\ttZbb}{\ttZ{}(\Zbb{})\xspace}
\newcommand{\ttZcc}{\ttZ{}(\Zcc{})\xspace}

\newcommand{\ttplus}[1]{\ensuremath{\ttbar{}+#1}}
\newcommand{\ttbb}{\ttplus{{\ge}2\PQb}\xspace}
\newcommand{\ttb}{\ttplus{\PQb}\xspace}
\newcommand{\ttlight}{\ttplus{\text{LF}}\xspace}
\newcommand{\ttcc}{\ttplus{{\ge}2\PQc}\xspace}
\newcommand{\ttc}{\ttplus{\PQc}\xspace}
\newcommand{\ttjets}{\ttplus{\text{jets}}\xspace}
\newcommand{\score}[1]{\ensuremath{\mathcal{D}_{#1}}}
\newcommand{\pNet}{\textsc{ParticleNet}\xspace}
\newcommand{\parT}{\textsc{ParT}\xspace}
\newcommand{\parTlong}{\textsc{Particle Transformer}\xspace}

\begin{document}

\begin{frontmatter}

\title{Search for \Hcc and measurement of \Hbb via \ttH production} 

\author[ugent]{Maarten De Coen}
\author{on behalf of the CMS Collaboration}
\affiliation[ugent]{organization={Universiteit Gent}, city={Ghent}, country={Belgium}}

\begin{abstract}
A search is presented for Higgs boson production in association with a top quark--antiquark pair (\ttH) with the Higgs boson decaying to a charm quark--antiquark pair (\Hcc{}). The same process with a Higgs boson decay to bottom quarks, \ttHbb, is measured simultaneously.  The analysis uses data from proton--proton collisions at 13\TeV collected with the CMS detector in 2016--2018, corresponding to an integrated luminosity of 138\fbinv. The observed \ttHbb{} cross section relative to its prediction from the standard model (SM) is \(0.91\,^{+0.26}_{-0.22}\). The observed (expected) results are compatible at the 95\% confidence level with a \ttHcc cross section that is at most 7.8 (8.7) times larger than the SM expectation. Assuming all other Higgs boson couplings to be SM-like, this sets an observed (expected) upper bound on the charm quark Yukawa coupling modifier of 3.0 (3.3) times its SM value. 

\end{abstract}

\begin{highlights}
\item Search for \Hcc{} in \ttH{} production using 13\TeV proton--proton collisions
\item Measurement of \ttHbb{} compatible with standard model (4.4$\sigma$)
\item Upper limit on \ttHcc{}  of 7.8 times the SM expectation (95\% CL)
 \item Upper limit on the charm Yukawa coupling modifier \(\kappa_c < 3\)
\end{highlights}


\end{frontmatter}

\section{Introduction}
\label{intro}

In 2012, the CMS~\cite{bib:CMS} and ATLAS~\cite{bib:ATLAS} Collaborations discovered a new particle with properties consistent with those of the Higgs boson as predicted by the standard model (SM). Since then, there have been many studies on the couplings between the Higgs boson and the fermions, known as Yukawa couplings~\cite{CMS:HIG-22-001,ATLAS:2022vkf}. The current focus lies on the second generation, where the charm quark has the largest Yukawa coupling. Upper limits on this coupling are available from searches in several Higgs boson production channels, e.g. in Refs.~\cite{ATLAS:2024yzu,CMS:HIG-21-008,CMS:2025qmm}.

In this contribution, we present a new channel where the Higgs boson is produced in association with a top quark--antiquark pair (\ttbar{}) and then decays into a charm quark--antiquark pair, \(\ttHcc\). We include a measurement of the process \ttHbb, and use \ttZbb and \ttZcc to validate our results. The dominant background is \ttbar{} with additional jet radiation, \ttjets{}. When those jets are charm- or bottom-flavoured, it is challenging to disentangle them from the signal processes. To do so, we use advanced machine learning techniques, both for jet flavour tagging as well as for event classification.

\section{Analysis strategy}
\label{analysis}

\begin{figure}[t]
	\centering
	\includegraphics[height=0.47\linewidth]{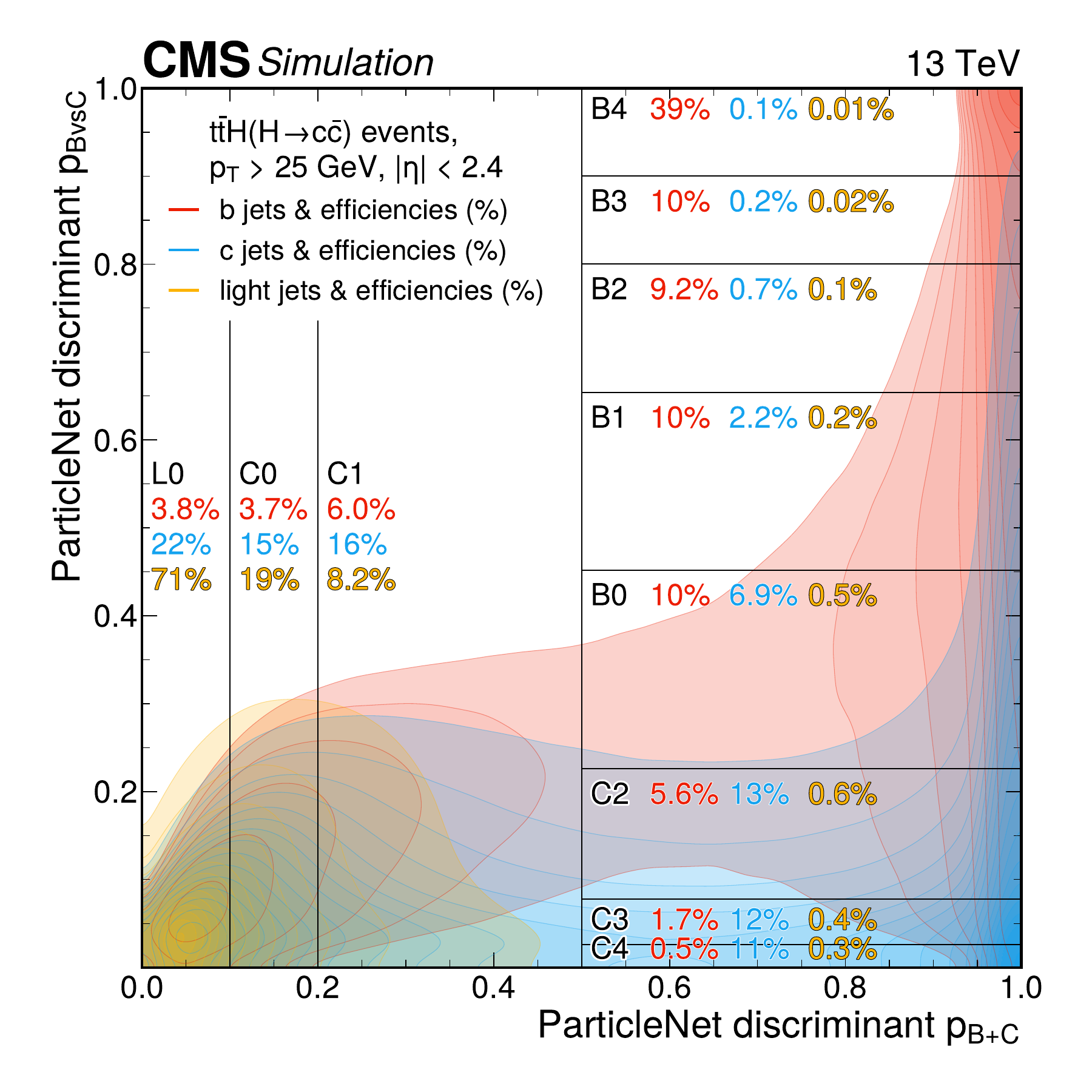}
	\caption{Jet flavour categories derived from two ParticleNet scores, separating charm- from bottom-flavoured jets (vertically) and heavy from light jets (horizontally). The contours represent the distributions of the differently flavoured jets in the plane.~\cite{CMS:2025dsh}}\label{partnet}
\end{figure}

The search is performed using data collected by the CMS detector between 2016 and 2018, for a total integrated luminosity of 138\fbinv{}. We consider all three decay channels of the \ttbar pair: fully hadronic (0L), semileptonic (1L), and dileptonic (2L). We select events with a high number of jets (7, 5, and 4 in the respective channels), minimal missing transverse momentum of 20\GeV in the 1L and 2L channels, and a sum of jet transverse momenta above 500\GeV in the 0L channel. Jet flavour tagging is performed using the graph-neural-network-based algorithm \pNet~\cite{Qu:2019gqs}. Based on its output scores, we categorize jets into light jets and several categories of charm- and bottom-flavoured jets, as shown in Fig.~\ref{partnet}. We require at least 3 jets tagged as \PQb jets (categories \texttt{B1}--\texttt{B4}, 70\% efficient) or \PQc jets (categories \texttt{C1}--\texttt{C4}, 50\% efficient), with at least 1 jet tagged as a \PQb jet.

The selected events are fed into a transformer-based neural network called \parTlong (\parT)~\cite{Qu:2022mxj}. Its inputs are properties and pairwise features of the event objects (jets, leptons, missing transverse momentum). The network is trained in each channel separately to assign events a likelihood score for the considered signal and background classes. The signal classes are \ttHcc, \ttHbb, \ttZcc, and \ttZbb. The \ttjets background is split into 5 classes, depending on the flavour of the extra jets. We assign \ttbar{} events with 1 (at least 2) additional b jet to the class \ttb (\ttbb). Otherwise, events with  1 (at least 2) additional c jets fall into \ttc (\ttcc). All remaining \ttbar{} events are classified as \ttlight{}. In the 0L channel, we include an additional background class of events composed exclusively of jets produced through quantum chromodynamics (QCD) interactions, or QCD multijet events for short. 

The \parT scores (denoted by \score{}) are used to further select and categorize events. We only consider events with a high sum of all signal scores \(\score{\ttX{}} > 0.6\) and with \(\score{\ttplus{LF}} < 0.05\) (0.02 in 1L). In the 0L channel, the QCD multijet background is reduced to negligible levels by selecting events with \(\score{QCD} < 10^{-4}\). After these selections, we assign events passing \(\score{\ttX{}} > 0.85\) to one of 4 signal regions (SRs), each enriched in one of the signal processes. All other events fall into one of 5 control regions (CRs) designed to constrain the normalizations of the five \ttjets{} backgrounds.
We then perform a binned profile likelihood fit to data. In each category, the fitted distribution is the corresponding \parT discriminant. For each signal process, we fit a signal strength \(\mu\): the ratio of the observed event yield to the SM expectation. The five \ttplus{\text{jets}} background normalizations are also left freely floating. Other backgrounds are modelled with simulation. Experimental and theoretical uncertainties are included in the fit as nuisance parameters.

\section{Results}
\label{results}

\begin{figure}
	\centering
	\includegraphics[width=0.55\linewidth]{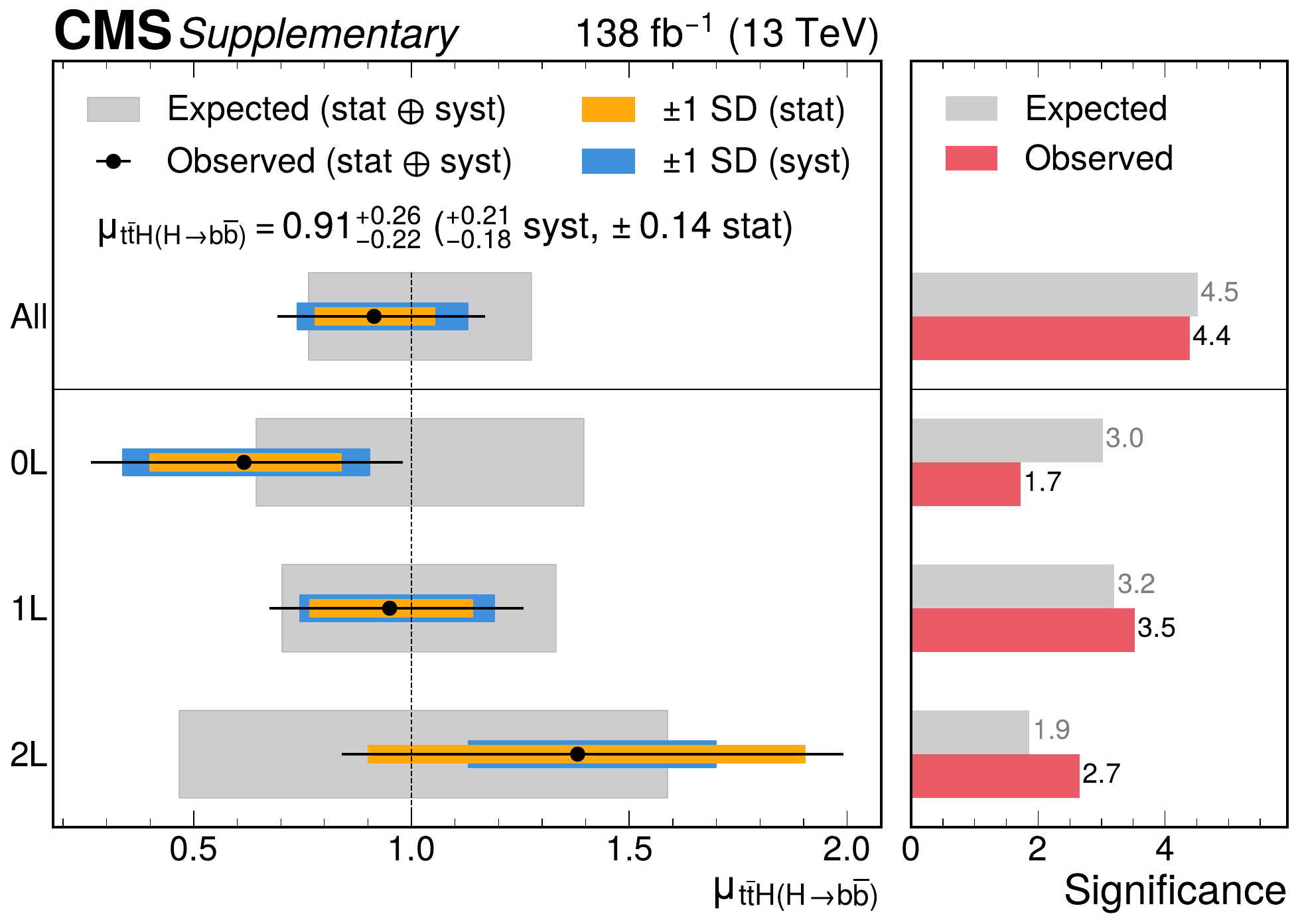}
	\includegraphics[width=0.35\linewidth]{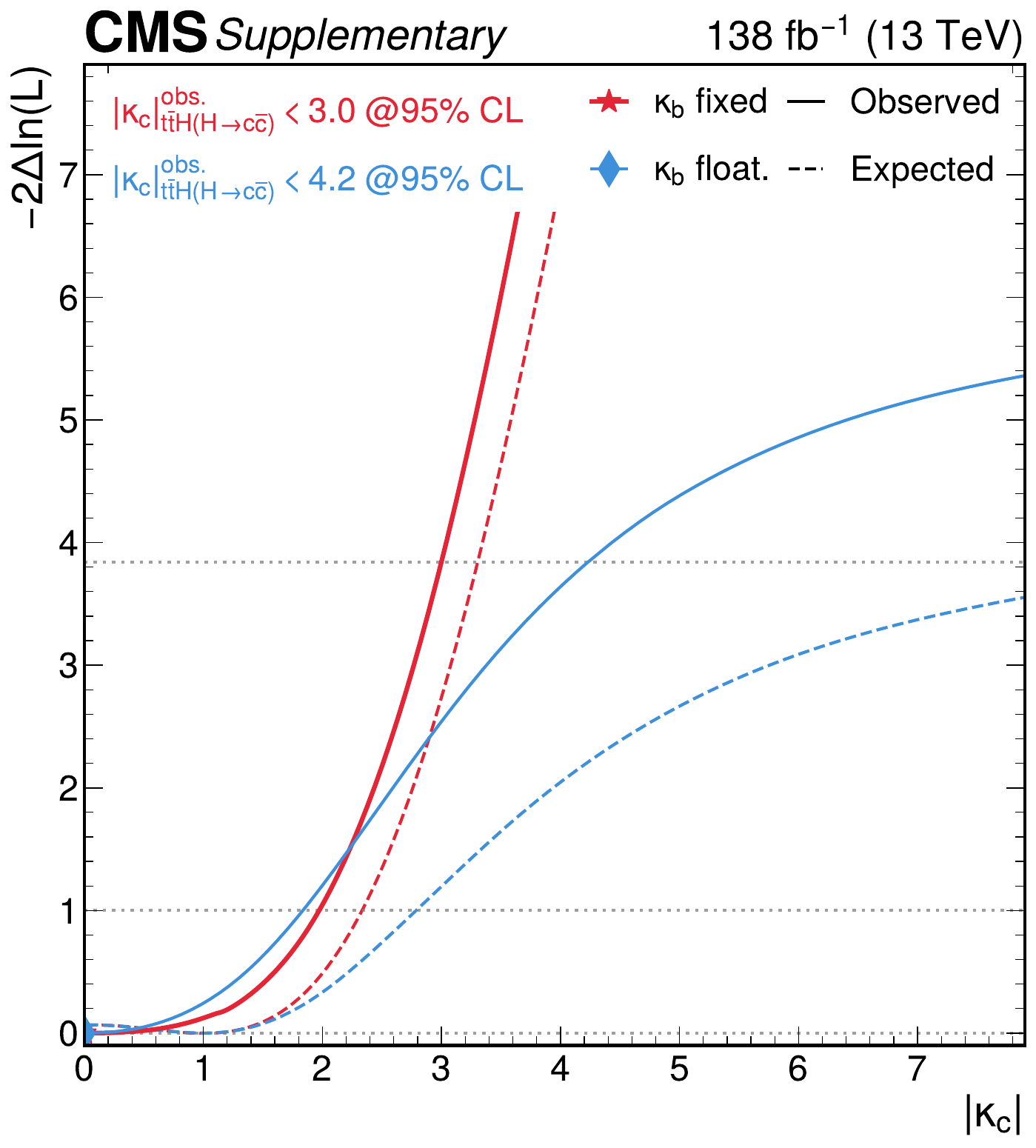}
	\caption{Left: measured and expected \ttHbb signal strengths in the three channels and combined. Right: likelihood scan of the Yukawa coupling modifier \(\kappa_c\) for fixed or floating \(\kappa_b\). ~\cite{CMS:2025dsh}}
	\label{fig:musigmatthbb}
\end{figure}

The measured (expected) signal strengths for \ttZcc and \ttZbb are \(\mu_{\ttZcc} = 1.02\,^{+0.79}_{-0.84}\) and \(\mu_{\ttZbb} = 1.47\,^{+0.45}_{-0.41}\). Both are compatible with SM expectations at the 95\% confidence level, as well as with previous measurements~\cite{CMS:2024mke,ATLAS:2023eld}. The measured \ttHbb signal strength is \(\mu_{\ttHbb} = 0.91\,^{+0.26}_{-0.22}\) which corresponds to 4.4 standard deviations above the background-only hypothesis. The contribution of each channel is shown in Fig.~\ref{fig:musigmatthbb}~(left). The dominant systematic uncertainty stems from the theoretical prediction of the \ttbb{} background. For \ttHcc{}, the measured (expected) limit on the signal strength is \(\mu_{\ttHcc} < 7.8\ (8.7)\) at the 95\% confidence level based on the CLs criterion. This result is statistically dominated, and will improve in the future with more data taking. The leading experimental uncertainty comes from the jet flavour tagging.
We interpret the measured signal strengths in the \(\kappa\)-framework~\cite{deFlorian:2016spz}, which parametrizes the branching ratios for \Hcc and \Hbb in terms of the Yukawa coupling modifiers \(\kappa_b\) and \(\kappa_c\). Fixing \(\kappa_b\), we find an (expected) limit of \(\kappa_c < 3\ (3.3)\) (see Fig.~\ref{fig:musigmatthbb}), which improves on the previous best limits set by CMS and ATLAS from Higgs boson production in association with a vector boson~\cite{CMS:HIG-21-008,ATLAS:2024yzu}. The combination of our result with the CMS result from this channel gives an observed (expected) limit of \(\kappa_c < 3.5\ (2.7)\).

\bibliographystyle{elsarticle-num} 
\bibliography{biblio}{}
\end{document}